\documentclass[%
 aip,
 amsmath,amssymb,
 reprint,%
]{revtex4-1}

\usepackage{graphicx}
\usepackage{dcolumn}
\usepackage{bm}

\usepackage[utf8]{inputenc}
\usepackage[T1]{fontenc}
\usepackage{mathptmx}
\usepackage{etoolbox}

\makeatletter
\def\@email#1#2{%
 \endgroup
 \patchcmd{\titleblock@produce}
  {\frontmatter@RRAPformat}
  {\frontmatter@RRAPformat{\produce@RRAP{*#1\href{mailto:#2}{#2}}}\frontmatter@RRAPformat}
  {}{}
}%
\makeatother
\begin{document}

\preprint{AIP/123-QED}

\title[ Microscopic flexoelectricity ...]{ Microscopic flexoelectricity in the canonical PMN relaxor  }

\author{J. Hlinka}
\thanks{Corresponding author. Email: hlinka@fzu.cz}
\affiliation{FZU - Institute of Physics of the Czech Academy of Sciences\\%
Na Slovance 2, 182 00 Prague 8, Czech Republic}
%


\begin{abstract}
Previously reported  neutron scattering investigations of the canonical relaxor ferroelectric perovskite oxide with the chemical formula Pb(Mg$_{1/3}$Nb$_{2/3}$)O$_{3}$ (PMN) are revisited to appreciate the role of the intrinsic bulk flexoelectricity.
Despite the outstanding electromechanical properties of lead-based relaxors, the magnitude of the flexoelectric coupling coefficient $f_{11} - f_{12}\approx -2V$, derived here directly from the PMN neutron diffuse scattering data, does not exceed the range of values typical for conventional perovskite ferroelectrics. 
We explain how these findings are related in the framework of the Ginzburg-Landau-Devonshire and the ferroelectric soft mode theory.
We propose that the relaxor properties of PMN might be related to the 
suppression of the  transverse correlation length of the flexoelectrically hybridized translational-polarization fluctuations due to its closeness to the Lifshitz-point regime. 
\end{abstract}

\maketitle

%


\section{Introduction}

Generation of electric polarization in an insulating solid by elastic strain gradients is a challenging engineering task.
The linear part of the underlying universal mechanism, known as flexoelectricity, as well as the associated converse effect, has already been the subject of numerous experimental and theoretical investigations, and it promises remarkable possibilities for the design  of unique electromechanical materials properties at the nanoscale.
At the microscopical level, it often plays the key role in theories of modulated ferroelectrics and antiferroelectrics\cite{Axe70, Etxebarria92, Pottker, Liubachko, Tagantsev13}.
Recent comprehensive overviews of the flexoelectricity with a focus on perovskite oxides can be found, for example, in Refs.\,\onlinecite{Yudin13,Wang19}. 

To support the current flexoelectric engineering efforts, it would be convenient to employ the most advanced computer-aided approaches, allowing to explore various combinations of nanoscale geometry and distinct materials. Obviously, for that, an extensive registry of materials properties is needed. However, although the flexoelectric effect appears to be easily accessible for direct measurements, in practice, these measurements are often difficult because of the sensitivity, nonlinearity, surface phenomena or other experimental issues. Potentially, the flexoelectric tensor describing the intrinsic bulk properties of crystals can be calculated ab-initio following the theory exposed e.g. in Refs.\,\cite{Hong13,Stengel}, but these techniques are still facing the computational limits. Consequently, flexoelectric tensor elements are so far established only for a rather limited number of materials\cite{Yudin13,Wang19}.

In this work, we address the intrinsic bulk flexoelectricity in one of the most remarkable perovskite oxide relaxor ferroelectric crystal, the cubic lead magnesium niobate, PMN. 
The magnitude of its flexoelectric property will be derived here from parameters of its characteristic atomistic structural inhomogeneity, that has been originally extracted  from the extensive experimental neutron scattering study of PMN diffuse scattering in Ref.\,\onlinecite{Vakhrushev}.
The essential ingredient of our analysis relies on the observation,  highlighted and emphasized in \onlinecite{Hirota02}, that the typical single unit cell pattern (eigenvector) of the frozen structural modulations differs from the eigenvector of the dynamic soft optic phonon by a systematic uniform shift,  oriented in the direction and in the positive sense of the local frozen polarization. 
We explain that this shift, called a "phase shift" there, can be understood as a natural consequence of the converse flexoelectricity, and we use it to estimate the magnitude of the flexoelectric coefficient   $f_{11} - f_{12}$, as well as the lower bound for the corresponding gradient coefficient $G_{11} - G_{12}$.
Moreover, we argue that the system is close to the Lifshitz point and that this proximity is one of the critical ingredients responsible for the polar nanodomain formation in perovskite relaxors.

\section{"Phase shift" model of Hirota et al.}

By inspecting neutron diffuse scattering intensities of the PMN relaxor single crystal in multiple inequivalent Brillouin zones, Vakhrushev et al. in Ref.\,\onlinecite{Vakhrushev} determined that there must be inhomogeneous distribution of atomic displacements superposed over its cubic average structure, which locally, {\it i.e.} at the unit cell level, consists
in the set of parallel atomic displacements $\{\delta_i\}$, $i=1-5$ with the following relative amplitudes:
\begin{eqnarray}
 \delta_1=\delta_{\rm Pb}&=&1.00,    \label{d} \\ \nonumber
    \delta_2=\delta_{\rm MN}&=&0.18,\\ \nonumber
    \delta_3=\delta_4=\delta_5=\delta_{\rm O}&=&-0.64 . \nonumber
\end{eqnarray}
These quantities are normalized by the lead ion displacement $d_{\rm abs}$. 
The lead displacements are disordered across the lattice, on average reaching $d_{\rm abs} \approx 0.25-0.3$\,\AA\, at the lowest temperatures. 
What matters in the following is that 
 these five values $\delta_i$ can be considered as a sort of universal "frozen phonon eigenvector" of PMN.

Hirota et al. in Ref.\,\onlinecite{Hirota02} pointed out  that
(i) this pattern does not  satisfy the center-of-mass condition for optic phonon modes, or in other words,
\begin{equation}
\sum_{\kappa} \delta_{\kappa} m_{\kappa} \neq 0, 
\end{equation}
where $m_{\kappa}$ is the mass of the atom $\kappa$ and the summation goes over the 5-atom perovskite unit cell; 
that (ii)  there is a shift $\tau\approx 0.58$ such that the "shifted eigenvector"
\begin{equation}
  h_{\kappa}=  \delta_{\kappa} -\tau
  \label{rozklad}
\end{equation}
would satisfy the center-of-mass condition, and
that (iii) this  "shifted eigenvector" 
\begin{eqnarray}
 h_1=h_{\rm Pb}&=&0.42, \label{h}\\   \nonumber 
    h_2=h_{\rm MN}&=&-0.40,\\  \nonumber 
    h_3-h_4=h_5=h_{\rm O}&=&-1.22   \nonumber 
\end{eqnarray} 
is  consistent with the dynamical structure factor of the soft phonon mode observed in their {\it inelastic} neutron scattering experiments.

In agreement with the intuitive picture of the frozen ferroelectric nanodomains or frozen polar nanoregions, the ultimately short range correlations should be of a ferroelectric type.
The diffuse scattering is sensitive to the correlations in the distribution of these frozen entities.
In particular, the characteristic  anisotropy of diffuse scattering in various Brillouin zones suggests that the dominant contribution can be described as a superposition of harmonic polarization waves in the form
\begin{equation}
   {\bf P}({\bf r}) = {\bf P}({\bf q}) \sin({\bf q}\cdot {\bf r} +\varphi),
   \label{Pvlna}
\end{equation}
where vectors ${\bf q}$ and  ${\bf P}({\bf q})$  fall along one of the twelve $\langle 110 \rangle$-type crystal directions, and at the same time, ${\bf q}$ and  ${\bf P}({\bf q})$ satisfy the transversality condition, that is,  
\begin{equation}
  {\bf P}({\bf q}) \in \langle 110 \rangle ~ \wedge ~
  {\bf q} \in \langle 110 \rangle ~ \wedge ~
  {\bf P}({\bf q}) \cdot {\bf q}=0 ,
  \label{hvezda}
\end{equation}
what is fulfilled for example with  ${\bf P}({\bf q}) \parallel$[1-10] and ${\bf q}\parallel$[110] and obviously all $m\bar{3}m$-symmetry equivalent pairs.
Since the range of $|{\bf q}|$ is broad, a plausible real-space pattern might contain bundles of differently spaced lamellar nanodomains  with $(110)$-type boundaries, at which the [1-10]-component of the frozen polarization fluctuations alternates its sign -- similarly as, for example, in the case of electromechanically compatible O180(110) or T90(110) walls of barium titanate\cite{Marton}. 

The condensation of ${\bf P}({\bf q}) \parallel$[1-10] and ${\bf q}\parallel$[110] waves is obviously in the heart of the so-called pancake model\cite{Xu2004}, but let us stress that the  assumptions of eqs.\,(\ref{Pvlna},\ref{hvezda}) about the frozen wave basis do not necessarily imply that the resulting local polarization vector has to preferentially point along one of the twelve $\langle 110 \rangle$ directions.
This can be illustrated with the help of the notation for ferroelectric domain walls, which explicitly includes direction of the domain-wall normal $|{\bf n}\rangle$ as well as directions of the spontaneous polarization in each of the two adjacent domains\cite{Goncalves2024}.
For example, the nominally uncharged R109(110) domain wall
of $[1\bar{1}1|110\rangle \bar{1}11]$ type can be obtained as a superposition of nominally uncharged O90(110) wall of $[1\bar{1}0|110\rangle \bar{1}10]$ type (which satisfies the above geometry requirements) with a homogeneous [001]-polarized background (which does not contribute to the diffuse scattering around the Bragg reflection position).
Thus, diffuse scattering by waves described by eqs.\,(\ref{Pvlna},\ref{hvezda}) is well compatible with the usual assumption that the resulting preferential local polarization favors $\langle 111 \rangle$ type crystallographic directions\cite{Pasciak2007, Pasciak2012,Eremenko2019}.
In fact, strictly speaking, the atomistic pattern of eq.\,(\ref{h}) might apply only to the correlated  fluctuations of the frozen polarization satisfying eq.\,(\ref{hvezda}), but for simplicity, let us assume that it applies to all components of the disordered polarization of PMN.

Summarizing, the model of Ref.\,\onlinecite{Hirota02} implies that each major frozen polarization wave ${\bf P}({\bf q})$ satisfying eq.\,(\ref{hvezda}) has a purely optical part associated  with the eigenvector of the soft optic phonon mode given by eq.\,(\ref{h}), and an ultimately related translational wave, given by  ${\bf u}({\bf q}) \propto {\bf P}({\bf q})$. 
The last relation has a common proportionality factor.
However, the reasons for this simple relationship have not been fully clarified there.

\section{Flexoelectric perspective}

Let us inspect these facts from the point of view of the Ginzburg-Landau-Devonshire theory of perovskite ferroelectrics. The integrand of the  Gibbs free energy functional expressed in terms of the Cartesian components $P_{i}$, $\epsilon_{ij}$ of the polarization and strain fields and their derivatives reads
\begin{eqnarray}
  \Delta G&=&  \frac{1} { 2} \alpha_{ij}P_{i}P_{j} + \frac{1} { 4} \beta_{ijkl}P_{i}P_{j}P_{k}P_{l} +
  \label{LGD}\\ \nonumber
 &+& \frac{1} { 2} G_{ijkl} P_{i,j} P_{k,l} +
 \frac{1} { 2} C_{ijkl} \epsilon_{ij} \epsilon_{kl}  - \\ \nonumber
  &-&  q_{ijkl} \epsilon_{ij} P_{k}P_{l} 
  - \frac{1} { 2} f_{ijkl}( \epsilon_{ij} P_{k,l}   - \epsilon_{ij,l}P_{k} )  - \\ \nonumber
  &-&P_{i}E_{i} + ... ~ ,
\end{eqnarray}
where the Einstein summation convention applies and the indices after the commas mean the partial derivatives. 
%
%
In what follows, the most important are the elastic, electrostrictive and flexoelectric coupling tensors $C_{ijkl}$, $q_{ijkl}$ and $f_{ijkl}$.
In particular, the third line of eq.\,(\ref{LGD}) also includes the "educated" sign convention in which the principal components of the electrostrictive and flexoelectric tensors for the known perovskite oxides are positive. 
Let us stress that  $f_{ijkl}$ should be distinguished from another flexoelectric tensor $\mu_{ijkl}^E =\epsilon_{im} f_{mjkl}$ that combines $f_{ijkl}$ coupling with the permittivity tensor $\epsilon_{im}$ and which is often introduced to describe the experiments in which the polarization is induced directly by the strain gradients at the fixed electric field\cite{Wang19,Narvaez,Ma}.

In agreement with the usual treatment of the strain as a  secondary order parameter, there are only terms linear and quadratic in strain fields in eq.\,(\ref{LGD}). Consequently, for a given polarization distribution and boundary conditions, there is a unique strain field that can, in principle, be calculated from the corresponding Euler-Lagrange equations\cite{Hlinka and Klotins}. The general solution is rather complicated even under periodic boundary conditions when the homogeneous strain can be conveniently treated separately. Nevertheless, in the short-wavelength limit, that is, for strongly inhomogeneous polarization and strain fields, the flexoelectric terms dominate over the electrostrictive ones, and the Euler-Lagrange equation reduces to\cite{Yudin13}
\begin{equation}
    C_{ijkl}\epsilon_{kl} \doteq f_{ijkl}P_{k,l},
\end{equation}
where $\epsilon_{kl} = ( u_{k,l} +u_{l,k})/2$.
For a given polarization sinusoidal wave ${\bf P}({\bf q})$ of a high symmetry, the acoustic  displacements ${\bf u}$ that satisfy this equation are also simple sinusoidal modes.
In particular, the equation implies that to be in the mechanical equilibrium, each purely polarization sinusoidal wave with ${\bf P}({\bf q}) \parallel$ [1-10] and ${\bf q}\parallel$ [110] requires a  purely acoustic sinusoidal mode with the same ${\bf q}$ and with ${\bf u}({\bf q}) \parallel {\bf P}({\bf q})$  such that one is simply proportional to the other via\cite{Yudin14}
\begin{equation}
    {\bf u}({\bf q}) = \frac{f_{11}-f_{12}}{C_{11}-C_{12}} {\bf P}({\bf q})  .
    \label{umera}
\end{equation}
In this way,  we see that flexoelectric mechanism explains the existence of the "phase shift", the parallelism ${\bf u}({\bf q}) \parallel {\bf P}({\bf q})$, the universality and uniqueness of the ratio $ |{\bf u}({\bf q})|/|{\bf P}({\bf q})|$, and it also relates the experimentally determined sense of ${\bf u}({\bf q})$ with the usual property of the flexoelectric tensor of polar perovskites oxides $f_{11}>f_{12}$.

\section{Magnitude of the flexoelectric coupling}

The decomposition of the displacement pattern $\{\delta_i\}$ of  eq.\,(\ref{d}) according to eq.\,(\ref{rozklad}) 
can be used to estimate $f_{11}-f_{12}$. 
We note that the "phase shift" $\tau$ is the displacement of the center of mass of the unit cell that is subject to the dimensionless displacement pattern $\{\delta_i\}$
\begin{equation}
  \tau=  \frac{\sum_{\kappa} \delta_{\kappa} m_{\kappa}}{\sum_{\kappa} m_{\kappa}}
  \label{tau}
\end{equation}
and the amplitude of the purely acoustic component in a cell with a real lead-ion displacement $d_{\rm abs}$ is $\tau d_{\rm abs}$.
The polarization of such an unit cell can be evaluated from the formula 
\begin{equation}
  p=  \frac{\sum_{\kappa} Z_{\kappa} \delta_{\kappa}}{V_{\rm cell}}d_{\rm abs} = \frac{\sum_{\kappa} Z_{\kappa} h_{\kappa}}{V_{\rm cell}} d_{\rm abs} .
\end{equation}
Insertion of the known estimates\cite{Bellaiche} of atomic Born effective charges $Z_{\kappa}$  of PMN and its unit cell volume $V_{\rm cell}\approx 65.9 $\AA$^3$ yields
$p/d_{\rm abs}\approx 26$\,GC/m$^3$ = 26\,GPa/V, and
\begin{equation}
\frac{|P(q)|}{|u(q)|} =\frac{p}{u} =  \frac{p}{\tau d_{\rm abs}} \approx 44.5 {\rm ~ GPa/V} .
\end{equation}
Using eq.\,(\ref{umera}) and elastic tensor values $C_{11}-C_{12} \approx 80$\,GPa from Ref.\,\onlinecite{Ahart07} or $C_{11}-C_{12} \approx 100$\,GPa from Ref.\,\onlinecite{Stock12} one can finally obtain the desired estimate
\begin{equation}
f_{11}-f_{12} =(2 \pm0.2)V~.
\label{odhad}
\end{equation}
This value is similar to those reported for other perovskite oxide ferroelectrics\cite{Wang19}.

The harmonic stability analysis within the above Ginzburg-Landau-Devonshire model of eq.\,(\ref{LGD}) results in upper bounds on the flexoelectric coefficients compatible with the ferroelectric instability, and in particular,\cite{Yudin14}
\begin{equation}
|f_{11}-f_{12} | \leq \sqrt{ (C_{11}-C_{12}).(G_{11}-G_{12})}.
\label{Lifshitz}
\end{equation}
The same condition implies an important lower bound on the gradient coefficients defining the acceptable stiffness of  the ${\bf P}({\bf q})$ polarization waves satisfying eq.\,(\ref{hvezda})
\begin{equation}
(G_{11}-G_{12}) \geq G_{\rm min},
\label{mez}
\end{equation}
where for the above numerical estimates for PMN, $ G_{\rm min}\approx 0.4$\,10$^{-10}$ Jm$^3$C$^{-2}$. 
Let us stress that the eq.\,(\ref{Lifshitz}) is not absolute physical limit on the possible values of flexoelectric coefficients, it holds only within the adopted truncation of the Landau-Ginzburg-Devonshire expansion and the assumed simple ferroelectric instability, nevertheless, the results of eq.\,(\ref{odhad}) and eq.\,(\ref{mez}) are robust and should be relevant for designing phase-field models for PMN relaxor.

\section{Lattice-dynamics perspective}

Let us now address the "phase shift" model  directly in terms of discrete lattice dynamics. Phonon theory for ABO$_3$ cubic perovskite would typically work with $15$-component  vectors $|v\rangle  =\{ \langle \kappa i|v\rangle\} = \{v_{\kappa i} ({\bf q})\}$, where $\kappa$ distinguish the 5 atomic sublattices per unit cell and the subscript $i$ specifies each Cartesian component of the dispacement. 
The  instantaneous displacement in a direction $i$ of an atom located at ${\bf R_{L}} + {\bf r}_{\kappa}$ can be defined as
\begin{equation}
 Re[v_{\kappa i} ({\bf q}) e^{i {\bf q} \cdot ({\bf R} + {\bf r}_{\kappa}) } ],   
\label{definice}
\end{equation}
where  ${\bf R_{L}}$ is the crystal lattice vector of the unit  cell and ${\bf r}_{\kappa}$ are suitably chosen relative positions of the atoms of the reference cell at the origin.

The eigenmodes of the dynamical matrix can be cast as orthonormal with respect
 to the mass-weighted scalar product and the related norm
\begin{equation}
\langle v|w\rangle \equiv \frac{\sum m_{\kappa }v_{\kappa i} w_{\kappa i}^*}{\sum m_{\kappa }}, 
\parallel v \parallel^2 \equiv \frac{\sum m_{\kappa }v_{\kappa i} v_{\kappa i}^*}{\sum m_{\kappa }}.
\label{product}
\end{equation}

The norm $\delta$ of the frozen displacement pattern of eq.\,{\ref{d}}  then reads
\begin{equation}
\delta = \sqrt{\frac{\sum_{i=1}^5 m_{\kappa } \delta_{\kappa i}^2}{\sum_{i=1}^5 m_{\kappa }}} \doteq 0.84,
\end{equation}
and the norm $h$ of its optic part (of the "shifted eigenvector") satisfies $h^2=\delta^2-\tau^2$, where the norm of the acoustic part $\tau$ is given by the eq.\,(\ref{tau}).

Within the ansatz of eq.\,(\ref{definice}), it is natural to use a given optical pattern to introduce the whole  branch of "bare soft optic phonons" $|{\rm TO}\rangle$, polarized along
${\bf e}_{1-10} = (1,-1,0)/\sqrt{2} $ and propagating with an arbitrary perpendicular wavevector ${\bf q}$
\begin{equation}
\langle \kappa i {\bf q}|{\rm TO}\rangle \equiv  \frac{h_{\kappa}}{h} {\bf e}_{1-10}. 
\label{definiceTO}
\end{equation}
Similarly, for any such ${\bf q}$, one can define the "bare acoustic branch" $|{\rm TA}\rangle$ polarized along
${\bf e}_{1-10}$  through
\begin{equation}
\langle \kappa i {\bf q}|{\rm TA}\rangle \equiv {\bf e}_{1-10},   
\label{definiceTA}
\end{equation}
Note that both branches have ${\bf q}$-independent eigenvectors that are normalized to 1, so that we can assume that the generic condensed phonon eigenvector causing diffuse scattering at ${\bf q} \parallel  [110]$ is proportional to
\begin{equation}
|{\delta}\rangle = h|{\rm TO}\rangle + \tau|{\rm TA}\rangle.   
\end{equation}

 Likewise, in the parent paraelectric phase, it can be assumed that the eigenvector of the lowest frequency phonon transverse branch $|\omega_{-}\rangle$ (  often called an acoustic one)  can be
expressed as a linear combination of the $|{\rm TO}\rangle$ and $|{\rm TA}\rangle$ modes of eqs.\,(\ref{definiceTO},\ref{definiceTA}). 
The 2x2 dynamical matrix describing the mixing of $|{\rm TO}\rangle$ and $|{\rm TA}\rangle$ modes for ${\bf q} \parallel  [110]$ in their own basis can be written in the form
\begin{equation}
D({\bf q}) \equiv \left(
\begin{array}{cc}
 \omega_{\rm TO}^2 ({\bf q})    &  \Delta({\bf q})^* \\
    \Delta({\bf q}) &  \omega_{\rm TA}^2({\bf q})
\end{array}\right),
\label{matice}
\end{equation}
where the coupling term $\Delta({\bf q})$ vanishes in the ${\bf q} \rightarrow 0$ limit since the bare modes are exact eigenvectors by construction there, and $\Delta({\bf q})$ is real due to symmetry reasons discussed elsewhere\cite{Axe70}.
In the following, let us assume that $|\omega_{-}\rangle$ is the normalized exact eigenvector of $D({\bf q})$, corresponding to its smaller eigenvalue $\omega_{-}^2$: 
\begin{equation}
\omega_-({\bf q})^2 =   \langle \omega_- |D({\bf q})| \omega_- \rangle.
\end{equation}

\section{Critical soft modes}

In the soft mode theory of phase transitions of second order, the eigenvectors of the condensed phonon modes are same as the paraelectric soft modes that are approaching the stability limit.
Since the diffuse scattering of PMN revealed frozen modes with ${\bf q} = (q,q,0)a^*$, where $a^*$ the reciprocal lattice vector $a^*=2\pi/a_0$, it can be anticipated that there is a
whole  $\omega_-({\bf q})$ branch of paraelectric soft modes with this property in PMN.
In the $q$-range the condensation occurs, we thus expect that
\begin{equation}
   \omega_-({\bf q}) \approx 0 
\end{equation}
which immediately implies
\begin{equation}
\frac{\langle{\rm TO}| \omega_-\rangle }{\langle{\rm TA}| \omega_-\rangle} = -\frac{ \omega_{\rm TA}^2 }{\Delta}=
-\frac{\Delta}{ \omega_{\rm TO}^2 }
\end{equation}
and also that 
 \begin{equation}
|\omega_-\rangle \propto |\delta\rangle,
\end{equation}
 which then together implies that
\begin{equation}
 \Delta({\bf q}) \approx - \frac{\tau}{h} \omega_{\rm TA }({\bf q})^2 .
 \label{DeltaGuess}
\end{equation}
and
\begin{equation}
\omega_{\rm TO }({\bf q})  \approx - \frac{\Delta({\bf q})}{\omega_{\rm TA }({\bf q})} = \frac{\tau}{h} \omega_{\rm TA }({\bf q}) .
\end{equation}

\section{Lifshitz regime}

A remarkable aspect of the diffuse scattering in PMN is that the $q$-range of condensed phonons is particularly broad.
Then, the natural question arises, what conditions would favor a  broad range of frozen mode wavevector moduli $|q|$?
An obvious obstacle for the condensation of a single $|q|$ in relaxor materials is the frozen disorder. 
But what else would promote the range of the frozen $q$-vectors? From the small-displacement lattice dynamics perspective, the optimal situation would arise if  $\omega_{-}({\bf q}) =  0 $,   or  equivalently, $\det D ({\bf q}) = 0 $,
would hold in a finite $q$-interval around the Brillouin zone center, or still differently,
\begin{equation}
  \Delta({\bf q}) = -\sqrt{{\omega_{\rm TA }({\bf q})} \omega_{\rm TO }({\bf q})} .
  \label{strong}
\end{equation}
Let us stress that $\det D ({\bf q}) $ is an even function of $q$ with $\det D ({\bf 0}) \equiv 0$.
A direct consequence of eq.\,(\ref{strong}) in a finite interval is that $\partial^2/\partial q^2(\det D)=0$, which implies that ${\omega_{\rm TO }({\bf 0})}=0$, and another consequence  (of validity of eq. (29) within a whole finite interval) is that
\begin{equation}
\frac{\partial^4}{\partial q^4}(\det D)=0 ,
\label{weak}
\end{equation}
at $q=0$, which corresponds to the equality in eq.\,(\ref{Lifshitz}).
In fact, in the absence of the frozen disorder, the eq.(\ref{weak}) defines the ultimate phase boundary for the continuous ferroelectric phase transition,
similarly as in case of eq.\,(\ref{Lifshitz}). 
In other words, only for $\frac{\partial^4}{\partial q^4}(\det D)>0$,
the warping of the polarization generates a positive energy penalty, and the condensed ferroelectric domain walls will necessarily form with a positive energy.
In contrast, when $\frac{\partial^4}{\partial q^4} (\det D)<0$, the spatial variation of the frozen hybridized polarization allows to gain energy with respect to the homogeneous ferroelectric state,  and the second-order instability would proceed to an antiferroelectric, commensurate or an incommensurate modulated phase.

There have been predictions considering $\frac{\partial^4}{\partial q^4}(\det D)<0$ can occur in some ambient pressure perovskite oxide ferroelectrics\cite{Yamada,Tagantsev13}.
However, to our best knowledge, there is no clear evidence for it; rather, the incommensurate or antiferroelectric phases in oxide perovskites seems to stabilized by other mechanisms compatible with $\frac{\partial^4}{\partial q^4}(\det D)>0$.
The main conjecture this work is that PMN might be remarkably close to the Lifshitz type point, 
and that it  plays a considerable role in formation of its polar nanodomain structure.

\section{Tight binding phonon dispersion model}

Let is illustrate the above discussion using a simple tight-binding model for hybridized TA-TO phonon branches. 
In fact, we can employ here the canonical simple model as outlined for example in Ref.\,\onlinecite{waterfall},
only to apply it to the dispersion along ${\bf q} = (q,q,0)a^*$ instead of  along ${\bf q} = (q,0,0)a^*$ direction.
The elements of the dynamical matrix $D$ in eq.\,(\ref{matice}) are then given  as 
\begin{equation}
   \omega_{\rm TA}^2(q) =A \sin^2(\pi q  ) ,
\end{equation}
\begin{equation}
   \omega_{\rm TO}^2(q) =c(T) + B\sin^2(\pi q  ) ,
   \label{SMdisp}
\end{equation}
\begin{equation}
   \Delta (q) =d\sin^2(\pi q  ) .
\end{equation}

\begin{figure*}[ht]
\centering
\includegraphics[width=.49\columnwidth]{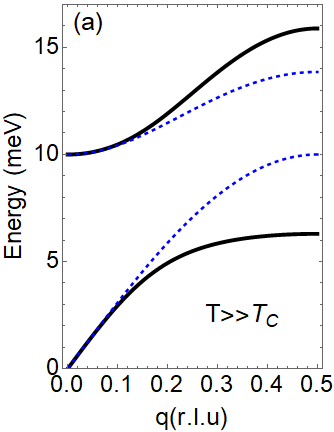}
\includegraphics[width=.49\columnwidth]{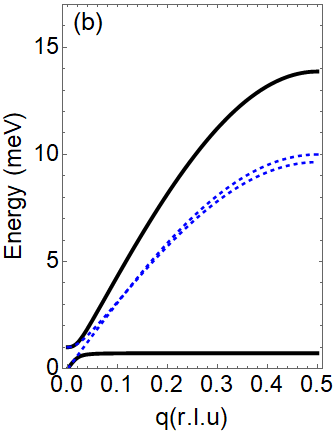}
\includegraphics[width=.49\columnwidth]{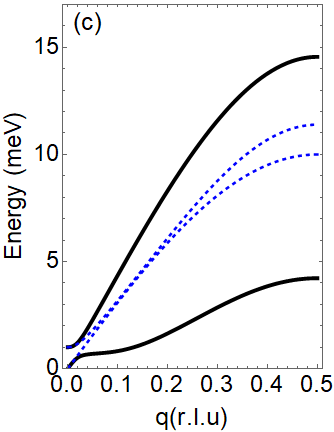}
\includegraphics[width=.49\columnwidth]{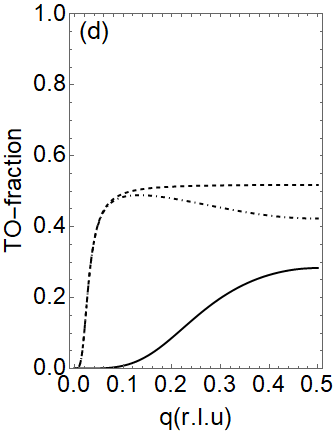}
\caption{Phonon dispersion curves in a model with flexoelectrically induced Lifshitz limit. 
(a) dispersion curves at temperatures well above the criticality, 
(b) the same but when the soft mode is approaching the phase transition,
(c) as previous, but with bare optic branch modified by admixing the second-neighbour interaction term. 
Dashed lines are bare TO and TA modes, full lines are eigenfrequencies of the dynamical matrix. 
The full, dotted and dash-dotted lines in the panel (d) gives the fraction of the bare TO mode $|\langle {\rm TO}| \omega_-\rangle|^2$ in the lower-frequency eigenmode $| \omega_-\rangle$, 
defining its "opticity" for  models of the panel (a),(b) and (c), respectively. Details are given in the main text\cite{ObligatoryZenodo}.
}
\label{fig:1}
\end{figure*}


The quasishear TA mode acoustic velocity is not renormalized by the coupling term and it corresponds to $A\approx 100$\,meV$^2$. 
Assuming that at the transition point the eq.\,(\ref{weak}) is fulfilled, we have $d=\sqrt{AB}$.
In order to reproduce the experimental $\tau/h$ ratio in the eq.\,(\ref{DeltaGuess}), we  set $B\approx 92$\,meV$^2$. 
The square of the TO mode frequency at the Brillouin zone center
$c\equiv \omega_{\rm TO}^2(0)$   is expected to be proportional to the $T-T_{\rm 0}$ temperature difference. 

At first sight, resulting dispersion curves calculated\cite{ObligatoryZenodo} in Fig.\,1a for $c=100$\,meV$^2$ appear rather usual and similar to the high-temperature inelastic neutron scattering results reproduced in Fig.\,16 of Ref.\,\onlinecite{Stock12}. 
The peculiarity of the parameters becomes apparent when the soft mode frequency is by one order of magnitude smaller (see Fig.\,1b.
There is a soft
branch with a very small frequency  along almost whole $\Gamma -M$ line of the Brillouin zone.
The reason behind this anomaly is that for $T=T_{\rm 0}$, within our simple model the weak condition of eq.\,(\ref{weak})  automatically implies the stronger requirement of eq.\,(\ref{strong}) along the whole $\Gamma -M$ line. 
In a more realistic description, there would be independent higher-order Fourier harmonics in the expansion of the all elements of the dynamical matrix, corresponding to the coupling between more distant atomic planes or among other lattice degrees of freedom. 
Example of such phonon branches is given in Fig.\,1c.
There we have replaced 40 percent of the soft mode dispersion in  eq.\,(\ref{SMdisp}) by the second-neighbor dispersive term $\sin^2(2\pi q  )$, while still keeping the constraint of eq.\,(\ref{weak})  and all other parameters as in Fig.\,1b.
We can see that there is still a significant portion of the anomalously flat dispersion near the Brillouin zone center, while near the Brillouin zone boundary,  the $| \omega_-\rangle$ branch regains the usual frequency range.
We speculate that this last case might at least roughly correspond to the experimental situation in PMN at temperatures below about 400\,K, where the long-wavelength infrared-active phonon mode critically slows down\cite{Zein,Hehlen}.

The admixing of the bare soft mode into the hybridized $| \omega_-\rangle$ branch  depends on both temperature and $q$, as illustrated in Fig.\,1d, but in the $c \rightarrow 0$ limit it should reach the limit value of
$A/(A+B) = (h/\delta)^2 \approx 0.52$ ratio for an arbitrarily low $q$ and, therefore,  the same limit ratio is expected to be transferred to all frozen modes.

Finally, Fig.\,2 shows the dispersion curves with the interaction term $d$ increased by 10 percent, that is, with the flexoelectric coupling increased beyond the "upper bound" given by the eq.\,(\ref{Lifshitz}). Except for the adjusted
temperature parameter $c$, all other parameters in Fig.\,2a and Fig.\,2b are the same as the parameters of Fig.\,1b and Fig.\,1c, respectively. The dispersion of the $| \omega_-\rangle$ branch  in Fig.\,2a and Fig.\,2b visibly corresponds to the phonon dispersions of a hypothetical system close to the second-order antiferroelectric and incommensurate instability, respectively. It also clarifies that the "upper bound" given by the eq.\,(\ref{Lifshitz}) can correspond to one of the two qualitatively distinct critical points, either to the paralectric--ferroelectric--antiferroelectric tripple point or to the paraelectric--ferroelectric--incommensurate Lifshitz point.

\begin{figure}[ht]
\centering
\includegraphics[width=.49\columnwidth]{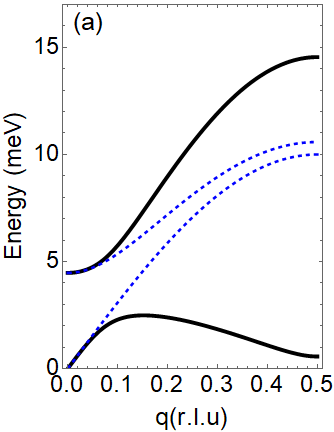}
\includegraphics[width=.49\columnwidth]{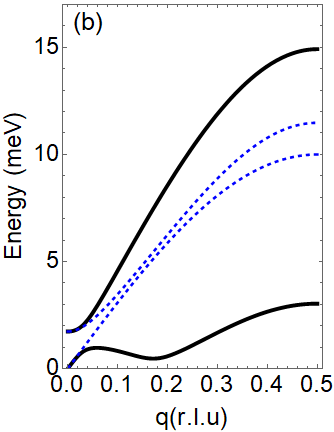}
\caption{Phonon dispersion with the interaction term $d$ increased by 10 percent with respect the Fig.\,1.
Panel (a) corresponds to a model with an antiferroelectric instability, $c= 20$\,meV$^2$, $d= - 105.5$\,meV$^2$, other parameters as in Fig.\,1b. Panel (b) corresponds to a model with an  incommensurate instability, $c= 3$\,meV$^2$, $d= -105.5$\,meV$^2$, other parameters as in Fig.\,1c.
}
\label{fig:2}
\end{figure}


\section{Discussion}

The fact that the flexoelectric coupling results in a renormalization of the correlation term $G_{ijkl}$ and that it does have an influence on domain patterns in ferroelectrics  is well known\cite{Ahluwalia2014}.
This influence of the flexoelectric coupling is expected to be most prominent when some eigenvalues of the  renormalized $G_{ijkl}$ tensor are pushed close to zero, that is, to the vicinity of the Lifshitz point. 

Obviously, so far the most attention was paid to materials and models in which an eigenvalue of the renormalized $G_{ijkl}$ becomes negative, so that it can drive incommensurate or antiferoelectric phases.

Nevertheless, a small positive eigenvalue of the renormalized $G_{ijkl}$ can play an important role as well.
In the mean-field approximation, the correlation lengths and domain wall widths scale with the square root of
the renormalized $G_{ijkl}$. 
For the renormalized transverse polarization fluctuations involved in the characteristic diffuse scattering of PMN (eq.\,\ref{Pvlna},\ref{hvezda}), the relevant correlation length $\xi$ should be sensitive to $G_{11}-G_{12}-G_{\rm min}$, 
and it can be expected that
\begin{equation}
\xi \approx \sqrt{ \frac{G_{11}-G_{12} -G_{\rm min}}{\alpha}  },
\label{delkaxi}
\end{equation}
where 
\begin{equation}
G_{\rm min} =   \frac{(f_{11}-f_{12} )^2}{C_{11}-C_{12}}  ,
\label{delkaxi}
\end{equation}
and $\alpha \approx |\alpha_{ii}|$ is the magnitude of the leading isotropic term of the Ginzburg-Landau expansion.

Although the correlation length and the width of the domain-wall in principle still diverge in the ideal limit $\alpha \rightarrow 0$, 
the actual magnitude of $\xi$ for any finite $\alpha$ should be effectively suppressed by the presence of $G_{\rm min}$ when $G_{11}-G_{12} \approx G_{\rm min} $. 
This circumstance, together with the frozen randomness related to the distribution of the Mg and Nb ions, could therefore favor the formation of the peculiar nanoscale polar domain texture with a significant domain wall pinning (as opposed to significantly broader domains and thicker mobile domain walls expected below the phase transition in the standard perovskite ferroelectrics).  

This situation, where $G_{11}-G_{12} - G_{\rm min} << G_{11}-G_{12}$, can be directly appreciated from the temperature-dependent dispersion curves illustrated in Fig.\,1. In particular, we can see that even when the bare optic branch is rather steep (far from $T_{\rm C}$, determined by $G_{11}-G_{12}$),  the lower frequency branch near $T_{\rm C}$ can in fact have a rather flat dispersion  (given by $G_{11}-G_{12} - G_{\rm min}$ see Fig.\,1c).
One possibility for experimental verification of these predictions would be a detailed comparative analysis of the phonon dispersion in the parent paraelectric phase by inelastic neutron data measured in different Brillouin zones. However, performing these experiments with the required high precision is still a challenge. Moreover, the essential information  about the phonon frequency dispersion is partially masked by a significant damping of soft modes in PMN.
An illustrative example of the scattering intensity calculated considering the same description of the damping sector of the phonon Green function model as in Ref.\,\onlinecite{waterfall}, when selecting the soft mode damping and the acoustic mode zone boundary damping to be 4\,meV, is shown in Fig.\,3. Clearly, the most interesting part of the soft mode dispersion is already in the fully overdamped regime where an independent extraction of the phonon frequency and damping becomes a difficult task.

\begin{figure}[ht]
\centering
\includegraphics[width=.89\columnwidth]{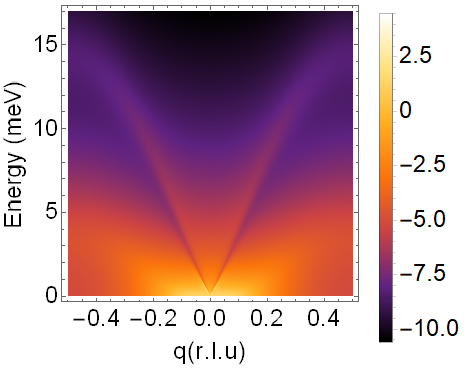}

\caption{Logarithm of the inelastic scattering intensity (in arbitrary units) by the coupled phonon dispersion of Fig.\,1c, calculated according the model of Ref.\,\onlinecite{waterfall},  when selecting the soft mode damping and acoustic mode zone boundary damping to be 4\,meV. The acoustic to optic scattering ratio was set to 1:4 in agreement with the left part of Fig.3 of Ref.\,\onlinecite{waterfall}.
}
\label{fig:3}
\end{figure}


Obviously, the ultimate quantitiative model for polar nanoregion formation
might potentially require not only the involvement of dynamics and disorder, but also some additional terms in the free energy expansion,
like the gradient electrostriction terms\cite{Hlinka and Klotins} or the invariants in the form of  $P_iP_jP_{k,l} P_{m,n}$. However, the elucidation of the process of formation of polar nanocluctures is a long-standing and difficult problem that is clearly far beyond the scope of the current work.

\section{Summary}

In summary, the Vakhrushev-Hirota model\cite{Vakhrushev,Hirota02} assuming a strict ratio between the amplitude of the frozen optic and acoustic modes in PMN can be understood from two different points of view.
The first one assumes the prevailing flexoelectric nature of the polarization-strain coupling, which is justified for large enough wavevectors.
The second one assumes that the frozen pattern closely follows the eigenvectors of the lowest frequency hybridized soft branch, which is expected to be justified if the mode amplitudes are small and independent.
Both approaches relate the observations to the $f_{11}-f_{12}$ combination of the components of the flexoelectric tensor.
The experimental results implies that $f_{11}-f_{12}\doteq (2 \pm0.2)V$, which is a fairly conservative result {\it per se}.
The lattice-dynamical approach also suggests that the gradient coefficient $G_{11}-G_{12}$ might be close to the Lifshitz limit.
The phonon dispersion curves corresponding to the limit case are analyzed in a simple model and estimated in Fig.\,1c.
It can be concluded that adjustment of gradient coefficients $G_{11}-G_{12}$ with respect to the flexoelectric and elastic properties of perovskite oxides could be a key element for the formulation and understanding of design rules for novel relaxor, incommensurate, and antiferroelectric perovskite oxides.

\begin{acknowledgments}
This work was supported by the Czech Science Foundation (project no. 25-15518L PFANDL) and by the Ferroic Multifunctionalities project, supported by the Ministry of Education, Youth, and Sports of the Czech Republic, Project No. CZ.02.01.01/00/22\underline{\,\,\,}008/0004591, co-funded by the European Union. It is a great pleasure to acknowledge the friendly support and useful critical comments by P. Ondrejkovi\v{c} from FZU.
\end{acknowledgments}

\end{document}